\documentclass[11pt]{article}
\usepackage{times} 
\usepackage{subfigure} 
\usepackage{amsmath} 
\usepackage{amsfonts} 
\usepackage{amsthm} 
\usepackage{graphicx}
\usepackage{latexsym}
\usepackage{fullpage}
\usepackage{delarray}
\usepackage{url}
\usepackage{verbatim}
\def\MyStretch{1.1}
\def\baselinestretch{\MyStretch}

\newcommand{\urlBiBTeX}[1]{\url{#1}} %%% needed for URL to print nice
\newcommand{\urlbibteX}[1]{\url{#1}} %%% needed for URL to print nice
\newtheorem{proposition}{Proposition}

\title{\bf 
Near rationality and 
competitive equilibria in networked systems\thanks{This work is
supported in part by the National Science Foundation through grants
ANI-0085879 and ANI-0331659.}}
\author{ 
{\em Nicolas Christin} \hspace{0.5cm} {\em Jens Grossklags} \hspace{0.5cm} {\em John Chuang}\\
School of Information Management and Systems\\ 
University of California, Berkeley\\
Berkeley, CA 94720\\
{\tt\{christin,jensg,chuang\}@sims.berkeley.edu}
}

\date{
{\small\em Technical Report, University of California, Berkeley}\\
\url{http://p2pecon.berkeley.edu/pub/TR-2004-04-CGC.pdf}\\
\vspace{0.5cm}
April 2004}

\begin{document}
\maketitle

\setcounter{footnote}{0}
\thispagestyle{empty}
\def\mynormality#1{\def\baselinestretch{#1}\small\normalsize}

\mynormality{1.0}

\vspace{0.5in}
\begin{abstract}
% $Id: abstract.tex,v 1.5 2004/04/19 18:24:57 christin Exp $
A growing body of literature in networked systems research relies on
game theory and mechanism design to model and address the potential lack
of cooperation between self-interested users. Most game-theoretic models
applied to system research only describe competitive equilibria in terms
of pure Nash equilibria, that is, a situation where the strategy of
each user is deterministic, and is her best response to the strategies
of all the other users. However, the assumptions necessary for a pure
Nash equilibrium to hold may be too stringent for practical systems.
Using three case studies on computer security, TCP congestion control,
and network formation, we outline the limits of game-theoretic models
relying on Nash equilibria, and we argue that considering competitive
equilibria of a more general form may help reconcile predictions from
game-theoretic models with empirically observed behavior.
\end{abstract}

\mynormality{1.2}

\newpage

% $Id: intro.tex,v 1.12 2004/04/19 18:24:58 christin Exp $ 
\section{Introduction}
\label{sec:intro}
Empirical evidence of phenomena such as free-riding in peer-to-peer
systems \cite{Adar:FirstMonday00} or unfairness in ad-hoc networks
\cite{Hsieh:SIGMETRICS01} challenges the traditional system design
assumption that all users of a network are able and willing to
cooperate for the greater good of the community. Hence, system
architects have become increasingly interested in considering
network participants as selfish \cite{Shneidman:IPTPS02} or
competing \cite{Shenker:TON95} entities. For instance, in an
effort to discourage free-riding, some deployed peer-to-peer
systems such as KaZaA or BitTorrent \cite{Cohen:p2pecon03} rely
on simple incentive mechanisms. More generally, as summarized in
\cite{FS:DAMD,Christos:STOC01,Shneidman:IPTPS02}, a number of recent
research efforts have been applying concepts from game theory and
mechanism design to networked systems in an effort to align the
incentives of each (self-interested) user with the goal of
maximizing the overall system performance.

A cornerstone of game theory and mechanism design is the notion of
competitive equilibrium, which is used to predict user behavior
and infer the outcome of a competitive game. As discussed in
\cite{Christos:STOC01}, the concept of {\em Nash equilibrium} is
predominantly used in system research to characterize
user behavior. Assuming each user obtains a utility dependent on
the strategy she adopts, a Nash equilibrium is defined as a set of
strategies from which no user willing to maximize her own utility has
any incentive to deviate \cite{Nash51}.

While Nash equilibria are a very powerful tool for predicting
outcomes in competitive environments, their application to system
design generally relies on a few assumptions, notably, that (1)
each participant is infallible (i.e., perfectly rational), and that
(2) each user has perfect knowledge of the structure of the game,
including strategies available to every other participant and their
associated utilities. There seems to be a class of problems for which
these assumptions may be too restrictive, for instance, characterizing
competitive equilibria in systems where participants have limited
knowledge of the state of the rest of the network.

As a practical example of the potential limits of a game theoretical
analysis of a networked system solely based on Nash equilibria, one can
argue that, in the case of a peer-to-peer file-sharing system that does
not provide incentives for users to share, the unique Nash equilibrium
leads to the ``tragedy of the commons \cite{Hardin68},'' that is, a
situation where users do not share anything to minimize the cost they
incur, thereby leading the entire system to collapse. The mere fact
that, in practice, some users are sharing files, even in peer-to-peer
systems that do not rely on incentive mechanisms, hints that a Nash
equilibrium is not actually reached.

In this paper, we argue that successfully applying game theory in
networked systems may require to consider competitive equilibria of a
more general form than pure Nash equilibria. We illustrate our point by
presenting three case studies, on security, TCP congestion control, and
network formation, where outcomes predicted by Nash equilibria are not
entirely correlated by empirical observations. In each case study, we
investigate if and how more general forms of competitive equilibria can
be used to better describe observed behavior.

The remainder of this paper is organized as follows. In
Section~\ref{sec:prelim}, we provide some background by formally
discussing the concepts of Nash equilibria and their extensions or
potential alternatives. In Section~\ref{sec:case}, we present our
case studies. Finally, in Section~\ref{sec:discussion}, we discuss
our findings, outline a possible agenda for future research, and draw
conclusions from our observations.

% $Id: prelim.tex,v 1.13 2004/04/20 05:29:54 christin Exp $
\section{Background}
\label{sec:prelim}

We consider strategic interactions (called {\em games}) of the
following simple form: the individual decision-makers (also called
{\em players}) of a game simultaneously choose actions that are
derived from their available strategies. The players will receive
payoffs that depend on the combination of the actions chosen by
each player.

More precisely, consider a set $N = \{1, ..., n\}$ of players. Denote as
$S_i$ the set of {\em pure} (i.e., deterministic) strategies available
to player $i$, and denote as $s_i$ an arbitrary member of $i$'s strategy set.
A probability distribution over pure strategies is called a {\em mixed}
strategy $\sigma_i$. Accordingly, the set of mixed strategies for
each player, $\Sigma_i$, contains the set of pure strategies, $S_i$,
as degenerate cases. Each player's randomization is statistically
independent of those of the other players. Then, $u_i$ represents player
$i$'s payoff (or {\em utility}) function: $u_i(\sigma_i, \sigma_{-i})$
is the payoff to player $i$ given her strategy $(\sigma_i)$ and the
other players' strategies $($summarized as $\sigma_{-i})$. An $n$-player
game can then be described as $G = \{N; \Sigma_i, \Sigma_{-i}; u_i,
u_{-i}\}$.

Players are in a Nash equilibrium if a change in strategies by any
one of them would lead that player to obtain a lower utility than
if she remained with her current strategy \cite{Nash51}. Formally,
we can define a Nash equilibrium as follows: {\em A vector of
mixed strategies $\sigma^{*} = (\sigma^{*}_1, ..., \sigma^{*}_n)
\in \Sigma$ comprises a mixed-strategy Nash equilibrium of a game
G if, for all $i \in N$ and for all $\sigma^{'}_i \in \Sigma_i$,
$u_i(\sigma^{'}_i, \sigma^{*}_{-i}) - u_i(\sigma^{*}_i,
\sigma^{*}_{-i}) \leq 0$.} A pure-strategy Nash equilibrium is a
vector of pure strategies, $s^{*} \in S$, that satisfies the
equivalent condition.

The economics community has provided an increasing number of refinements
to strengthen the concept of the Nash equilibrium, for example, to
remove counter-intuitive or unrealistic predictions. Complementary to
these refinements some have investigated the rational choice assumptions
on which the Nash equilibrium concept is built. For instance, a rational
player is expected to demonstrate error-free decision-making, to have
perfect foresight of the game and to be unbounded in her computational
abilities. Intuitively, players such as network users or automated
agents will likely deviate from these rigid assumptions.

Consider, for example, an experienced player whose strategy choice
is almost perfectly correlated with a Nash prediction of a game
but always contains a small error. She is playing in an auction
with an asymmetry between the expected cost of overshooting and
undershooting the Nash solution. If overshooting is less costly, 
the player's strategy will most likely contain a small upward
bias. If a substantial part of the other players shares this
marginal bias the outcome of the auction can be surprisingly far
away from a Nash prediction \cite{Goeree01}. Similarly, in a
sealed-bid auction the Nash equilibrium outcome predicts that a
player with a lower valuation will only sometimes win the
auctioned good. However, this outcome is more likely if players
share little imperfections in the execution of Nash strategies
\cite{Klemperer03}.

Such systematic and non-systematic deviations and their outcomes have
been motivation to formulate more generalized models of strategic
behavior that include the notion of the Nash equilibrium as a special
case. Examples are models that introduce (possibly small) amounts of
noise into the decision-making process \cite{Goeree04,McKelvey95}.
These models are very useful as an empirical structure for uncovering
features of payoffs from field data, or to obtain relationships between
observables and primitives of interest \cite{Haile03}. Another set of models derive equilibria that are {\em near 
rational}
\cite{Akerlof85,Radner80}. In near rational equilibria a player who
is not perfectly maximizing her utility cannot improve her payoff by
a substantial amount by playing her Nash strategy more accurately.
While the personal losses for a player are potentially very small, 
the equilibria derived often represent substantial departures from a
prediction based on perfect Nash optimizing behavior. These models are
appropriate for the description of empirical phenomena but can also
contribute explanations and predictions of strategic behavior. 

In the analysis we present in this paper, we will focus on a
simple, but powerful model of near rationality, called  the {\em
$\varepsilon$-equilibrium} \cite{Radner80}. We point out that 
other equilibrium concepts can also be useful in modeling and
analyzing networked systems, but defer the analysis of their
applicability to future work.

\begin{comment}
Strategies of near rational players and the derived outcomes can capture
many sources of deviations from a Nash prediction and can be computed
relatively easy given an existing analysis of rational best-response
equilibria.
\end{comment}

%\paragraph{$\varepsilon$-equilibrium}

The $\varepsilon$-equilibrium concept \cite{Radner80} is relaxing the
conception of a fully rational player to a model where each player
is satisfied to get close to (but does not necessarily achieve) her
best response to the other player's strategies. No player can increase
her utility by more than $\varepsilon$ by choosing another strategy.
Therefore, we locate an $\varepsilon$-equilibrium by identifying a
strategy for each player so that her payoff is within $\varepsilon$
of the maximum possible payoff given the other players' strategies.

Formally, an $\varepsilon$-equilibrium can be defined as follows:
{\em A vector of mixed strategies $\sigma^{\varepsilon} =
(\sigma^{\varepsilon}_1, ..., \sigma^{\varepsilon}_n) \in \Sigma$
comprises a mixed-strategy $\varepsilon$-equilibrium of a game G if,
for all $i \in N$, for all $\sigma^{'}_i \in \Sigma_i$, and a fixed
$\varepsilon > 0$, $u_i(\sigma^{'}_i, \sigma^{\varepsilon}_{-i}) -
u_i(\sigma^{\varepsilon}_i, \sigma^{\varepsilon}_{-i}) \leq \varepsilon$.}
A pure-strategy $\varepsilon$-equilibrium is a vector of pure strategies,
$s^{\varepsilon} \in S$, that satisfies the equivalent condition. For
$\varepsilon = 0$ this condition reduces to the special case of a Nash
equilibrium. Thus, one can consider $\varepsilon$-equilibria as a more
generalized solution concept for competitive equilibria.

% $Id: case.tex,v 1.18 2004/04/20 05:29:54 christin Exp $
\section{Case studies}
\label{sec:case}
In this section, we present three case studies on security, TCP
congestion control, and network formation. For each of the case studies,
we describe the interaction between the different participants in terms
of a game. We then note the discrepancies between the game outcome as
predicted by a Nash equilibrium and the behavior observed empirically,
and discuss if more general forms of equilibria can lead to more
accurate predictions.

\subsection{Protection against security threats}
For our first case study, we look at the level of security users choose
in a network subject to a security threat. Specifically, we focus on
protection against potential distributed denial of service (DDoS)
attacks. In the first stage of a DDoS attack, an attacker looks for
a (set of) machine(s) whose control they can easily seize, to use as
a platform to launch an attack of larger magnitude. For instance, by
obtaining total control of a machine on a network, an attacker may be
able to retrieve passwords and gain access to more secure machines on
the same network.

We model here a network of $n$~users, who are all potential targets in
the initial stage of a DDoS attack. If we characterize the level of
computer security that each user~$i$ adopts by a variable $s_i$, the
user(s) with the lowest $s_i$ (i.e., $s_i = s_{\min} = \min_{i}\{s_i\}$)
will be compromised. We assume that each user can infer the security
level $s_i$ used by every other user (e.g., by probing), and no finite
security level $s_i$ can be selected to guarantee a protection against
all attacks. We further assume that the cost of implementing a security
policy $s_i$ is a monotonic increasing function of $s_i$. Specifically,
to simplify the notations, we consider here that each user~$i$ that is
not compromised pays $s_i$ to implement their security policy. The
compromised user(s), say user~$j$, pays a fixed penalty $P \geq s_i$
(for any $s_i$), independent of the security level~$s_{\min}$ she has
chosen.

While very simplified, we conjecture this game is a relatively accurate
model of the first stage of DDoS attacks that have been carried out in
practice \cite{Dittrich:trinoo}.\footnote{While this type of attack
shares some similarities with worm propagation, notably searching for
insecure machines \cite{Moore:Slammer}, a worm typically propagates
by infecting all machines on a network that are below a certain, {\em
fixed}, security level, which is different from our hypothesis that only
the machines with the lowest level of security are compromised.} We
defer the study of the deployment of the attack beyond the first stage
to future work.

\begin{proposition}
The game described above has a unique pure Nash equilibrium, where
all users choose an identical security level $s_i = P$.
\label{prop:security-nash}
\end{proposition}
Proposition~\ref{prop:security-nash}, whose proof we derive in 
Appendix~\ref{sec:appendix}, tells us that, for a Nash equilibrium
to hold, all users have to choose the highest level of security
available. However, available data from large networks, e.g.,
\cite{Cisco:Vulnerability}, documents that different systems present
highly heterogeneous security vulnerabilities, which in turn indicates
that implemented security levels are highly disparate across machines.
Hence, in the context of the security game we just described, a Nash
equilibrium does not seem to accurately describe observed behavior.

Some of the possible explanations for the heterogeneity of the
implemented security levels can be captured by more elaborate
equilibrium models. In particular, (1) users have incomplete information
on the levels of security deployed by other users, (2) the {\em
perceived} benefit of installing security patches may be smaller than
the overhead patching incurs, and (3) some users may be gambling
(knowingly or not) on the seriousness of the security threats they
face. These three arguments all make the case for considering 
$\varepsilon$-equilibria with mixed strategies, rather than a pure Nash
equilibrium.\footnote{One could also consider pure
$\varepsilon$-equilibria, but it can be shown that, for this specific
game, pure $\varepsilon$-equilibria produce results very close to
Proposition~\ref{prop:security-nash}.}

\begin{proposition}
There exist mixed-strategy $\varepsilon$-equilibria
with $\varepsilon \leq P/4$ where all chosen security levels are distributed
over the interval $[0, P]$.
\label{prop:security-mixed}
\end{proposition}
Proposition~\ref{prop:security-mixed}, which we prove in
Appendix~\ref{sec:appendix}, indicates that considering
$\varepsilon$-equilibria with mixed strategies allows us to predict
large dispersion of the chosen security levels, even for relatively low
values of $\varepsilon$. This result seems to be more in line with the
available measurement data. We further note that analogous results have
been recently derived to quantitatively model price dispersion phenomena
\cite{BM:RAND}, where assuming a Nash equilibrium likewise fails to
corroborate empirical measurements.

One can direct two critiques at the discussion on the security game
we just presented. First, the discrepancies between the behavior
predicted by a Nash equilibrium and that observed in practice may be
due to an inaccurate game model, rather than from assuming a specific
type of equilibrium. Second, one can argue that while the assumption
of perfect rationality, as required in a pure Nash equilibrium, is
very debatable when strategies are selected by humans (such as in the
security game), perfect rationality is a much more reasonable assumption
in the case of automated agents. We attempt to address these concerns by
discussing additional case studies in the remainder of this paper.

\subsection{TCP congestion control}
The second case study relies on a game-theoretic analysis of the TCP
transport protocol \cite{Akella02}. Each TCP sender relies on an
additive-increase-multiplicative-decrease (AIMD) algorithm to adjust its
sending rate in function of the congestion experienced on the path from
sender to receiver.

In \cite{Akella02}, Akella et al. present a game-theoretic analysis to
model competition between different TCP senders for three of the most
popular variants of TCP, namely, TCP Tahoe, TCP Reno and TCP SACK.
In the {\em TCP Game} they describe, players are the TCP sources ($i
\in \{1,\ldots,n\}$), which are allowed to adjust their individual
additive increase ($\alpha_i$) and multiplicative decrease ($\beta_i$)
parameters. In the TCP Game, the utility of each player is equal to her
goodput, which is defined as the total amount of data transfered over
a time interval, minus the amount of data that had to be retransmitted
(presumably because of losses in the network) over the same time
interval.

One of the insights presented in \cite{Akella02} is that, for TCP
SACK, a pure Nash equilibrium results in $\alpha_i \rightarrow \infty$
(infinite additive increase) if $\beta_i$ is held fixed, while $\beta_i
\rightarrow 1$ (no multiplicative decrease) if $\alpha_i$ is held fixed.
Simply stated, if all players in a TCP SACK network were behaving
according to a Nash equilibrium, they would simply turn off congestion
control, which would likely result in the network suffering from
complete congestion collapse. However, TCP SACK is increasingly deployed
on the Internet \cite{Allman00:CCR}, and yet, we do not
observe congestion collapse phenomena due to misbehaving TCP
sources.\footnote{In fact, the authors of \cite{Akella02} point out
that the Nash equilibria for TCP NewReno and TCP SACK are similar. TCP
NewReno and TCP SACK combined currently account for an overwhelming
majority of all traffic on the Internet, which hints that the observed
stable operation of the Internet probably does not result from having a
mix of different TCP variants in the network.}

One of the possible reasons proposed by the authors of \cite{Akella02}
for the continued stable operation of the Internet is that a given user
may face technical difficulties to modify the behavior of her machine to
behave greedily. We submit this potential explanation can be partially
captured by considering an $\varepsilon$-equilibrium instead of a Nash
equilibrium. The cost of modifying the behavior of a given machine can
indeed be viewed as a switching cost, to be included in the factor
$\varepsilon$.

For simplicity, we assume here that players can only modify their
additive increase parameter $\alpha_i$. (An analogous study can
be carried out if we allow changes to $\beta_i$.) The authors of
\cite{Akella02} show that, with TCP SACK, player~$i$'s utility (goodput)
is given by
$$
u_i(\alpha_i, \alpha_{-i}) = c\frac{\alpha_i}{A+\alpha_i} \ ,
$$
where $c$ denotes the total capacity (bandwidth-delay product divided
by the round-trip-time) of the bottleneck link, and $A=\sum_{j \neq i}
\alpha_j$. Therefore, having an $\varepsilon$-equilibrium implies
that, for any $\alpha'_i$, $u_i(\alpha'_i, \alpha_{-i})-u_i(\alpha_i,
\alpha_{-i}) \leq \varepsilon$, so that
\begin{equation}
c\frac{A(\alpha'_i-\alpha_i)}{(A+\alpha'_i)(A+\alpha_i)} \leq \varepsilon .
\label{eq:sack-alpha}
\end{equation}
If we allow $\alpha_i = 0$ and $\alpha'_i \rightarrow \infty$, an
$\varepsilon$-equilibrium can only occur for $\varepsilon \geq c$, that
is, when $\varepsilon$ is larger than the maximum utility achievable.
In such a scenario, $\varepsilon$ is so large that all players select a
value for their parameter $\alpha_i$ at random.

Adding the assumption that variations of $\alpha_i$ are bounded leads
to much more interesting results.\footnote{Note that there are several
possible justifications for bounding the variations on $\alpha_i$.
For instance, because obtaining perfect knowledge of the state of the
entire network is difficult (or impossible) for a given user, each user
may instead incrementally probe the network to discover her optimal
setting for $\alpha_i$. Such a probing behavior can be captured as a
repeated game where, for each repetition, $\alpha'_i-\alpha_i \leq
K$.} Specifically, let us impose $\alpha'_i-\alpha_i \leq K$ for
$K \in \mathbb{N}$. For simplicity, let us set the initial values
for $\alpha_i$ to the default value in TCP implementations, that
is, $\alpha_i = 1$ for all~$i$. Then, we have $A=n-1$ and $0\leq
\alpha'_i \leq K+1$. Substituting in Eq.~(\ref{eq:sack-alpha}), we have a
$\varepsilon$-equilibrium as soon as
$$ 
\varepsilon \geq c\frac{K}{n} \ . 
$$
Hence, in a network with a large number of TCP senders, the default TCP
implementation can be an $\varepsilon$-equilibrium for small values
of $\varepsilon$. This is one of the possible explanations why the
predicted Nash behavior that users would turn off TCP congestion control
primitives is not fulfilled.

\subsection{Network formation}
For our third case study, we briefly discuss network formation
by self-interested parties. Following seminal work in economics
\cite{JackWol96}, network formation has lately received relatively
significant attention in the networking research community.
We refer the interested reader to recent studies, such as
\cite{Chun:INFOCOM04,Fabrikant:PODC03}, for an in-depth discussion of the
problem, and only focus here on the potential limitations of using Nash
equilibria in the context of network formation.

We define a network as a set of $n$~nodes connected by a set of
$k$~directed links (where $k \leq 2n(n-1)$). Each node is used to store 
items that are of
interest to other nodes. We follow the generic network model described
in \cite{ChCh:IPTPS04} where each node can request items, serve items,
or forward requests between other nodes. As in \cite{ChCh:IPTPS04},
we assume shortest-path routing. Using a few simplifying assumptions
(e.g., all nodes are considered to have the same capabilities, all links
have the same establishment cost, and requests for items are uniformly
distributed over the entire network), the authors of \cite{ChCh:IPTPS04}
express the cost associated to each node~$i$ as
$$
C_i = \frac{s}{n}+lEd_{i,j}+rEb_{j,k}(i)+m\deg(i)\ ,
$$
where 
$Ed_{i,j}$ is the expected value of the topological distance (hop-count)
between node~$i$ and another node~$j$, $Eb_{j,k}(i)$ is the expected
value of the probability that node $i$ is on the path between two
arbitrary nodes~$j$ and $k$, and $\deg(i)$ is the out-degree of
node~$i$, that is, the number of nodes node~$i$ links to. The constants
$s$, $l$, $r$ and $m$ represent the nominal costs associated with
storing an item, retrieving an item one hop away, routing a request
between two other nodes, and maintaining a connection to another node,
respectively. From this cost model, we can immediately define the
utility of node~$i$, $u_i$, as 
\begin{equation}
u_i = -C_i \ . 
\label{eq:util-netform}
\end{equation}

Assume that nodes can choose which links they maintain, but do not have
any control over the items they hold, and honor all routing requests.
In other words, nodes are selfish when it comes to link establishment,
but are obedient once links are established.

\begin{proposition}
With the utility function given in Eq.~(\ref{eq:util-netform}), if $m
< l/n$, the fully connected network
where each node links to every other node is a unique pure Nash equilibrium. 
\label{prop:netform-nash1}
\end{proposition}
\begin{proposition}
If $m > l/n$, the star-shaped
network, where all links connect to or from a central node, is a pure Nash
equilibrium.
\label{prop:netform-nash2}
\end{proposition}

Propositions~\ref{prop:netform-nash1} and \ref{prop:netform-nash2},
whose proofs are in Appendix~\ref{sec:netform}, tell us that, if
maintaining links is cheap, or if the network is small, the only Nash
equilibrium is the fully connected network. If maintaining links is
more expensive, or if the network is large, a star-shaped network is
a possible Nash equilibrium.\footnote{In the limit case where $m$ is
exactly equal to $l/n$, any network constitutes a Nash equilibrium.}
While the star may not be a unique Nash equilibrium, the high aggregate
utility of the star \cite{ChCh:IPTPS04} suggests it may dominate other
potential Nash equilibria. We note that the authors of \cite{JackWol96}
obtain comparable results using a slightly different cost model.

Thus, we would expect predominance of fully-connected or star-shaped
networks in practice. While these types of topologies can indeed be
found in existing networks (e.g., many small local area networks use
star topologies), measurement studies of Internet topologies exhibit
much more varied results \cite{Faloutsos:SIGCOMM99}. Among the reasons
why Internet topologies do not solely consist of an interconnection
of star-shaped and fully connected networks, one can cite capacity
constraints \cite{Chun:INFOCOM04} or monetary incentives.

While proposing a game-theoretic model that accurately captures 
these additional factors is outside of the scope of this paper, we
simply point out that, if instead of considering Nash equilibrium,
we consider an $\varepsilon$-equilibrium, then, for any $m \in
[l/n-\varepsilon,l/n+\varepsilon]$, {\em any} network topology
constitutes an $\varepsilon$-equilibrium. (This can be proven
by simply including $\varepsilon$ in all the derivations of
Appendix~\ref{sec:netform}.) Additionally, if, to account for failures
in link establishment due for instance to lossy channels, we allow
nodes to use mixed strategies instead of being restricted to pure
strategies, we conjecture that the range of possible values for $m$ such
that any network is an $\varepsilon$-equilibrium is much larger than
$2\varepsilon$.

The outcome of this third case study is that allowing small deviations
from Nash equilibria can result in obtaining very different network
topologies at the equilibrium. This is something a network designer may
want to keep in mind if her objective is to have self-interested nodes
form a particular topology.

\section{Discussion}
\label{sec:discussion}

We have shown through case studies that considering competitive
equilibria of a more general form than pure Nash equilibria can be
beneficial in systems research. In particular, we discussed how
allowing players to slightly deviate from their optimal utility
can help reconcile game-theoretic models and observed player
behavior.

We note that, even in games for which a pure Nash equilibrium is
undesirable from the system designer's perspective, near rational
players may actually settle for a desirable outcome. This is a possible
explanation why the Internet does not suffer from congestion collapse,
despite the inefficiency of the Nash equilibrium in the TCP SACK
game. Conversely, potentially desirable outcomes associated with a
Nash equilibrium may prove difficult to reach unless all players
are perfectly rational. The security game we described presents an
instance of such a phenomenon. Thus, it appears that taking into account
uncertainty factors can be useful in both game specification and
mechanism design.

% Potential tradeoff between Complete model and Approximate, simpler
% model but with imperfections
An alternative to modeling near rationality is to consider fully
specified games, which capture all factors with any conceivable
influence on the game outcome. However, we argue that the two
approaches are not exclusive. In fact, refinements to the game
description are probably of interest when the near rationality assumption
yields substantial deviations from the outcome predicted by a Nash
equilibrium. Research on bounded-reasoning and bounded-optimality models
\cite{Russell95} provides a solid framework for such refinements.

As a follow-up on our case studies, we are interested in gathering
experimental data, through user surveys, on how security levels
are chosen in practice, and in investigating how well this data
can be described using game-theoretic models. We are also planning
on conducting simulation studies to assess the actual impact of
uncertainties and of mixed strategies on network formation.

Last, we believe that this research has uncovered a few open
problems that may warrant future investigation. First, our case
studies seem to show that considering other types of equilibria
besides Nash equilibria can help expand the applicability of
game-theoretic models to networked systems. While the
$\varepsilon$-equilibrium used in this paper is an interesting
tool, many other equilibrium models have been investigated in the
literature, e.g., \cite{Akerlof85,Goeree04,McKelvey95,Radner80}.
We conjecture that different types of equilibrium may be
appropriate for different networking problems, and believe that
providing a classification of networking problems according to the
specific types of equilibrium that best characterize them would be
valuable.

More generally, one can also ask how a game-theoretic model can capture
that the rationality of each participant may vary across users: some
users may be obedient, some others may be fully rational, some may be
faulty \cite{FS:DAMD}. Finding if and how game-theoretical models can
accommodate for heterogeneous populations of players may help us design
better systems, and certainly poses a number of interesting research
questions.

% $Id: appendix.tex,v 1.6 2004/04/19 18:53:16 christin Exp $ 
\appendix
\section{Proofs of Propositions~\ref{prop:security-nash} and \ref{prop:security-mixed}}
\label{sec:appendix}
We first consider that users are only allowed pure strategies, and prove
Proposition~\ref{prop:security-nash}. \begin{proof}[Proof of
Proposition~\ref{prop:security-nash}] Without loss of generality, we
assume that users $\{1, \ldots, k\}$, with $1 \leq k \leq n$, choose
a security level $s_{\min} < s_i$ for all $i \in \{k+1, \ldots, n\}$.
Thus, each user~$i$ for $i \in \{1,\ldots, k\}$ is compromised, and has
a utility $u_i = - P$. Users in $i \in \{k+1, \ldots, n\}$ cannot be
compromised because $s_i > s_{\min}$ and therefore have a utility $u_i =
- s_i$.

Suppose a user~$i$ in $\{1,\ldots, k\}$ were to increase her security
level to $s_i = s_{\min}+h$ for $h > 0$. User~$i$'s utility would become
$-s_{\min}-h$. However, because the original constellation of security
levels forms a Nash equilibrium, we know that such a change of strategy
results in a decrease of user~$i$'s utility for any $h > 0$. That is,
for any $h > 0$,
$$
-s_{\min}-h+P \leq 0 \ ,
$$
which reduces to $s_{\min} \geq P - h$ for any $h > 0$, so that
$s_{\min} \geq P$ by continuity. By hypothesis, $s_{\min} \leq P$, which
implies that $s_{\min} = P$. Since for any $i$, $s_{\min} \leq s_{i}
\leq P$, we obtain $k = n$, and, for any~$i$, $s_i = P$ is the only
possible Nash equilibrium. The utility of each user is $u_i = -P$,
and cannot be increased by picking a different security level, which
confirms that $s_i = P$ for all~$i$ constitutes a Nash equilibrium.
\end{proof}

Suppose now that users choose their security level probabilistically.
More precisely, the probability that user~$i$ picks a security level
$s_i$ below a value $s$ is characterized by the cumulative distribution
function (c.d.f.) $F_{s_i}(s) = Pr [s_i \leq s]$.

\begin{proof}[Proof of Proposition~\ref{prop:security-mixed}]
Consider the following continuous c.d.f. $F_{s_i}(s)$:
\begin{equation}
F_{s_i}(s) = \left\{
\begin{array}{ll}
0 & \mbox{if $s \leq 0$,} \\
1-\left(1-\frac{s}{P}\right)^{\frac{1}{n-1}} & \mbox{if $0\leq s < P$,}\\
1 & \mbox {if $s \geq P$.}
\end{array}
\right .
\label{eq:cdf}
\end{equation}
We use $Eu_i(s)$ to denote the expected value of the utility $u_i(s)$
in function of a security level $s$. Because $u_i(s) = -P$
if all users $j \neq i$ choose security levels higher than $s$, and
$u_i(s) = -s$ otherwise, we have
$$
Eu_i(s) = -P(Pr[s_j > s])^{n-1} - s(1-(Pr[s_j > s])^{n-1}) \ , 
$$
which can be expressed in terms of $F_{s_i}(s)$ as
\begin{equation}
Eu_i(s) = -P(1-F_{s_i}(s))^{n-1} - s(1-(1-F_{s_i}(s))^{n-1})  \ .
\label{eq:eui}
\end{equation}
Substituting $F_{s_i}(s)$ by its expression given 
in Eq.~(\ref{eq:cdf}), Eq.~(\ref{eq:eui}) reduces to 
$$
Eu_i(s) = -P+s\left(1-\frac{s}{P}\right) . 
$$
A study of the variations of $Eu_i(s)$ in function of $s \in [0, P]$
indicates that $Eu_i(s) \geq Eu_i(0) = -P$ and that $Eu_i(s) \leq
Eu_i(P/2) = -3P/4$. Thus, if we have $\varepsilon = P/4$, any variation
of the expected utility is smaller $\varepsilon$, which characterizes an
$\varepsilon$-equilibrium. In other words, we have shown, by providing a
specific c.d.f. $F_{s_i}(s)$, that there exist $\varepsilon$-equilibria
with $\varepsilon \leq P/4$ where the security levels $s_i$ can be
spread out over the entire interval $[0,P]$. Note that we only present
an existence proof here. It is unclear whether the chosen c.d.f
$F_{s_i}(s)$ is an accurate depiction of how security levels are chosen
in reality, and it is likewise entirely possible that there exist
other distributions of the security levels over $[0,P]$ that result in
$\varepsilon$-equilibria for $\varepsilon \ll P/4$. 
\end{proof}

% $Id: netform.tex,v 1.3 2004/04/19 18:53:16 christin Exp $ 
\section{Proofs of Propositions~\ref{prop:netform-nash1} and ~\ref{prop:netform-nash2}}
\label{sec:netform}
Here, we first show that the fully connected network is the only Nash   
equilibrium if and only if $m < l/n$, before showing that, if $m >   
l/n$, the star-shaped network characterizes a Nash equilibrium.         

\begin{proof}[Proof of Proposition~\ref{prop:netform-nash1}]
In a fully connected network, no node can create additional links. If
a given node~$i$ removes one of its links, $\deg(i)$ decreases from
$(n-1)$ to $(n-2)$, but, at the same time, one of the nodes $i'\neq i$ is
now at a distance of 2 from $i$. Thus, $Ed_{i,j}$ increases from 1 to
$$
Ed_{i,j}=\frac{n-1}{n}+\frac{2}{n}=1+\frac{1}{n} \ ,
$$
and the difference in utility for node~$i$, between the strategy of
removing one link and the strategy consisting in maintaining all links,
is $m-l/n$. To have a pure Nash equilibrium, we therefore need to have $m-l/n
\leq 0$, which is true if and only if $m \leq l/n$. 

Suppose now that we have $m < l/n$, and a network that is not fully
connected. In particular, consider that a node~$i$ can decide whether
to create a link to to another node~$i' \neq i$. Before addition of
the link $i \rightarrow i'$, $i'$ is at a distance $2 \leq d_{i,i'}
\leq n-1$ of $i$. After creation of the link $i \rightarrow i'$, $i'$
is at a distance 1 of $i$. Thus, by creating the link $i\rightarrow
i'$, $Ed_{i,j}$ at least decreases by $(2-1)/n = 1/n$. Adding the link
$i\rightarrow i'$ also results in $\deg(i)$ increasing by one, so that
that the addition of the link $i \rightarrow i'$ eventually results in
a change in the node~$i$'s utility equal to $-m+l/n$, which, by
hypothesis, is strictly positive. Hence, node~$i$ always has an incentive
to add links to nodes it is not connected to. Using the same reasoning
for all nodes, we conclude that the fully connected network is the
unique Nash equilibrium if $m < l/n$. 
\end{proof}

Consider now a star-shaped network, where all links connect to or from
a central node, say node 0, and assume that $m > l/n$. 

\begin{proof}[Proof of Proposition~\ref{prop:netform-nash2}]
Node~0 is fully connected to the rest of the network, and therefore
cannot create additional links. If node~0 removes one of its links, one
of the $n-1$ other nodes becomes unreachable, which implies $Ed_{0,j}
\rightarrow \infty$, and $u_0 \rightarrow -\infty$. Thus, node~0 has no
incentive in modifying its set of links. Likewise, peripheral nodes do
not remove their (only) link to the central node, to avoid having their
utility $u_i \rightarrow -\infty$.

Suppose now that a peripheral node~$i$ creates an additional link to
another peripheral node~$i' \neq i$. An argument identical to that
used in the proof of Proposition~\ref{prop:netform-nash1} shows that
the addition of the link $i \rightarrow i'$ results in a change in
the node~$i$'s utility equal to $-m+l/n$. Here, however, $m > l/n$,
so that $-m+l/n < 0$, and node~$i$ has no incentive in adding the
link $i\rightarrow i'$. Thus, the star-shaped network is a pure Nash
equilibrium, which may not be unique.
\end{proof}

\end{document}